\documentclass[aps,pra,reprint,groupedaddress]{revtex4-1}

\usepackage{graphicx}
\usepackage{amsmath}
\usepackage{float}

\begin{document}


\title{Capillary Assembly of Colloids: Interactions on Planar and Curved Interfaces}

\author{Iris B. Liu}
\altaffiliation
{Department of Chemical and Biomolecular Engineering, University of Pennsylvania, Philadelphia}

\author{Nima Sharifi-Mood}
\altaffiliation
{Department of Chemical and Biomolecular Engineering, University of Pennsylvania, Philadelphia}

\author{Kathleen J. Stebe}
\altaffiliation
{Department of Chemical and Biomolecular Engineering, University of Pennsylvania, Philadelphia}
\email{kstebe@seas.upenn.edu}


\date{\today}
\begin{abstract}
In directed assembly, small building blocks are assembled into an organized structures under the influence of guiding fields. Capillary interactions provide a versatile route for structure formation. Colloids adsorbed on fluid interfaces distort the interface, which creates an associated energy field. When neighboring distortions overlap, colloids interact to minimize interfacial area. Contact line pinning, particle shape and surface chemistry play important roles in structure formation.  Interface curvature acts like an external field; particles migrate and assemble in patterns dictated by curvature gradients. We review basic analysis and recent findings in this rapidly evolving literature. Understanding the roles of assembly is essential for tuning the mechanical, physical, and optical properties of the structure.
\end{abstract}

\pacs{surface tension, anisotropic colloids, Janus particles, curvature, self-organization, packing, inclusions, electrostatics }

\maketitle


\section{Introduction}
Over the past decade, the advent of complex colloids as building blocks has fueled intense interest in their organization and assembly to form new materials \cite{Glotzer}.  Anisotropic colloids with non-spherical shapes or patchy surfaces offer important degrees of freedom, including complex, directionally dependent potentials \cite{Glotzer, Romano}. Sophisticated assembly schemes include design of particle shape and chemistry to favor the formation of particular superstructures. For example, colloid shape can be designed to assemble into complex crystalline structures via emergent interactions owing to features like sharp edges and planar facets \cite{Damasceno, Teich}, or to interact  specifically via lock-and-key interactions \cite{SacannaNature}. Colloidal surface chemistry can be tailored, for example, via grafting of DNA  \cite{DNAcoated, Rogers2016,Valignat, Eiser,Kim} or tailored wetting \cite{PawarLang2008, PawarLang2009, BJP2012SM, BJP2013SM} to drive particular structure formation. These  examples exploit weak, $O(k_BT)$, interactions to direct assembly. In other modes of directed assembly, applied  electro-magnetic fields guide assembly. Particle interactions with the field and each other are typically very strong, and the resulting structures are often trapped. Particles or their assemblies orient and move along field lines to form superstructures with symmetries coupled to the field itself. Examples include para- or ferromagnetic colloids chaining along magnetic fields gradients \cite{BiswalPRE, BiswalPRE2003, LiuSM2015} and dipolar particles chaining in electric field gradients \cite{FurstPRE2000, Dommersnes}. In our research, we focus on fields that can direct colloid assembly because they rely on energy landscapes in confined soft matter. Examples include interface-shape mediated capillary energy fields for particles on fluid interfaces \cite{Cavallaro27122011}, membrane-shape mediated fields for colloids on lipid bilayer membranes \cite{Ningwei}, and elastic energy fields for colloids in confined nematic liquid crystals \cite{LuoSM2016, IrisPNAS, CavallaroLCPNAS, LeeAM2016}. These fields are remarkably versatile in their ability to guide microparticles into well defined structures. Furthermore, because they depend on the soft matter's configuration, which can be dynamically tuned, they are routes to reconfigurable structures.  \\
\indent In this review we discuss capillary interactions between microparticles at fluid interfaces. Classically, colloids trapped at fluid interfaces have been exploited to stabilize emulsions and foams \cite{Pickering}. The colloids can bring added functionality, for example, as catalysts \cite{Crossley2010} or as responsive structures to dynamically (de)stabilize emulsions \cite{Tu2014JACS}. Particles at interfaces can form disordered structures \cite{YunkerPRL} or ordered monolayers \cite{Pieranski}. For ordered systems, interface curvature imposes topological constraints \cite{Hall2016, Dinsmore1006, Irvine2010}. Electrostatic interactions between particles are known to play important roles in structure formation at interfaces, as has been widely discussed \cite{Leunissen, LeunissenNature, AveyardPRL2002, Ghezzi}. There are several excellent reviews that address these topics, which are outside the scope of this review \cite{McGorty2010MT, Kralchevsky2000145}.  \\ 
\indent At the scale of hundreds of microns to millimeters, there are many familiar examples of capillary interactions. In nature, "water striders" exploit surface tension to support their weight, and propel themselves using hydrophobic legs \cite{Hu2003}. The oft-cited Cheerios effect is another example, in which pieces of breakfast cereal cluster on the surface of a bowl of milk \cite{cheerios}.  The heavy pieces distort the interface around them, and interact to lower the interfacial and gravitational potential energies \cite{Chan, LeeVella1}, although details in the interface deformation around the morsels of cereal probably play a role in the near field. By dynamically tuning the interface around them, whirligig beetles raft to form chains and other structures \cite{Voise1357}, and waterlily leaf beetles larvae move along curved  menisci \cite{Hu}. These examples show that capillary interactions are highly shape dependent. The concept of designing particle shape to interact specifically was explored in an exciting body of work at this scale decades ago \cite{Bowden1997, Bowden1999JACS, Bowden2001}. Particles were designed with faceted shapes; some facets were well wet, others were not. At fluid interfaces, to minimize the excess area, they assembled to bring  well-wet facets in contact, and poorly wet facets in contact, to form well defined structures. \\ 
\indent In this review, we focus on capillary interactions as a means of microparticle assembly. We focus on interactions between colloids up to tens of microns, including disks, spheres, cylinders and ellipsoids. When microparticles are introduced to the interface by spreading or sedimentation from a dense suspension, they can become trapped in disordered layers owing to the strength of near field capillary interactions between them (colloidal monolayer membrane). However, if particles are sparse, or are introduced by sequential addition, they can assemble in preferred configurations to form a range of structures. Furthermore, at this length scale, interface curvature acts like  an external field; particles move along curvature gradient lines to particular sites and  form structures related to the underlying curvature gradients. Thus, at the microscale, complex structures can be formed even from simply shaped particles without complex, tailored wetting conditions. \\
\indent In this review we describe key concepts in the underlying physics of particles at fluid interfaces, drawing in part on notes prepared for the summer school 2015  International School of Physics "Enrico Fermi"\cite{Varenna}.  We review recent advances in the literature, and identify open issues and areas of ongoing research in the field. 
\section{SIMPLIFICATIONS OWING TO PARTICLE SIZE}
Capillary interactions occur between microparticles trapped at fluid interfaces as they move to minimize the interfacial area. In the event of contact line motion, particles move to minimize the sum of the energies owing to interface area and wetting energies. These interactions are determined by the wetting configuration of the particle and the shape of the interface around the particle. Typically, because of their microscopic radii $a$, several forces or stresses can be neglected, simplifying analysis of the particle interaction.  We enumerate several of these effects here. \\

(i) Particle weight or buoyancy can typically be neglected, with the Bond number:
\begin{eqnarray}
Bo = {{\rho g{a^2}} \mathord{\left/
 {\vphantom {{\rho g{a^2}} \gamma }} \right.
 \kern-\nulldelimiterspace} \gamma } \ll 1,
\end{eqnarray}
where $\gamma$ is the surface tension, $\Delta \rho$ is the difference in fluid densities, and $g$ is the acceleration due to gravity.  \\

(ii) Particle inertia can typically be neglected. Once attached to fluid interfaces, particles typically  move in creeping flow, with Reynolds number:
\begin{eqnarray}
Re={\rho U a}/{\mu} \ll 1,
\end{eqnarray}
where $\mu$ is a characteristics viscosity of the fluids near the interface and $U$ is the characteristic particle velocity. In this case, the sum of forces on the particles is zero.  \\

(iii) The interface shape is typically independent of particle velocity. The magnitude of viscous stresses compared to surface tension is negligible, with capillary number:
\begin{eqnarray}
Ca={\mu U}/{\gamma}\ll 1.
\end{eqnarray}  
This allows quasi-static analysis;  at any instant in time, capillary interactions are determined by the interface shape, which is determined only by the particle locations and contact line configurations. \\

(iv) Capillary interactions are often so strong that Brownian effects are negligible, with the P\'{e}clet number,
\begin{eqnarray}
Pe={Ua}/D\ll 1.
\end{eqnarray}  
where D is the Stokes-Einstein diffusivity of the particle in the interface, and U is the characteristic velocity of particle migration. In this limit, in creeping flow, capillary forces are balanced by viscous drag. This equality allows particle paths to be analyzed to find energy lost to viscous dissipation, and to infer capillary energy landscapes along those paths.\\

(v) Particles can deform the interface, with distortions that decay over distances comparable to the particle radius.  In analysis, the  height $h$ of the interface around the particle above a reference plane tangent to the interface is often described in a Monge gauge, i.e., interface height $h(\textbf{r})$ is a single valued function, where \textbf{r} is a position vector on the interface. Interface slopes are often assumed to be small compared to unity. In this limit, the shape of the interface is governed by
\begin{eqnarray}
{\nabla ^2}h = \frac{{\Delta P}}{\gamma }.
\end{eqnarray}  
where ${\Delta P}$ is the pressure difference evaluated at the interface. For constant mean curvature interfaces, the interface obeys: 
\begin{eqnarray}
{\nabla ^2}h = 0.
\end{eqnarray}  

There is, however, a body of work for particles on curved interfaces in which the particle-sourced distortions are assumed to decay slowly compared to the radius of curvature \cite{ZengSM2012, Guzowski}, or to move to finite slope \cite{Blanc2013PRL}.  We do not address those limits here.
  
\indent In the following sections, we discuss theory for particles at interfaces and review key findings. Since we focus on the roles of interface shape and wetting configurations, the theory applies equally to interfaces between immiscible fluids. We discuss, in turn, isolated particles, pair interactions and particles at curved fluid interfaces. 
\section{Trapping of isolated particles on planar interfaces} 
\begin{figure}[h]
\includegraphics[scale=0.6]{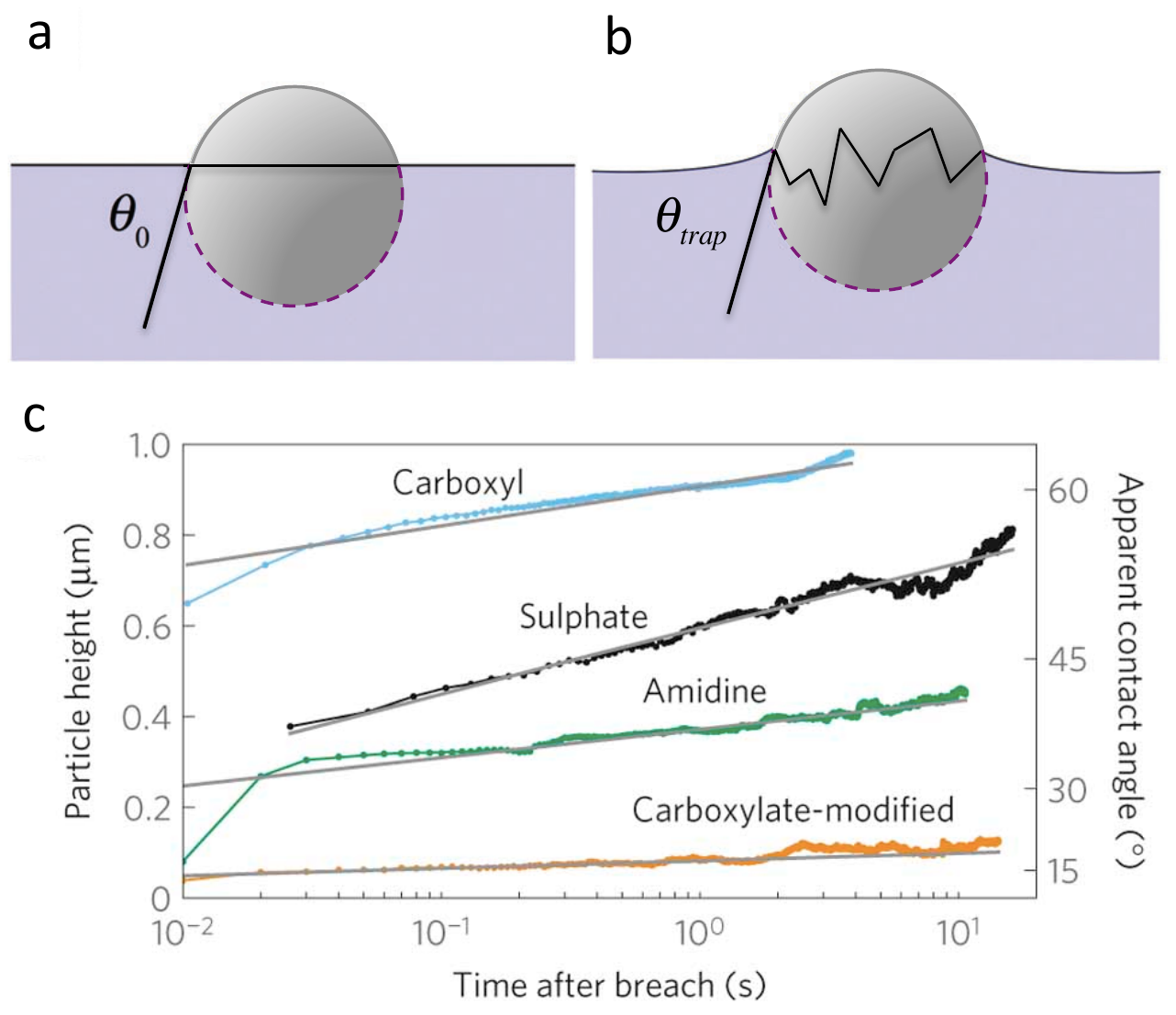}
\caption{Particle trapped at a fluid interface. Schematic of particle adsorption on an interface with (a) an equilibrium contact angle, and (b) pinned contact line. The gray solid lines represent the solid-vapor interfaces and the purple dashed lines represent the solid-liquid interfaces. (c) Glassy pinned contact line dynamics for different surface functional groups. Reprinted by permission from Macmillan Publishers Ltd: \textit{Nature Materials}, Reference \cite{Kaz:2012pin}, copyright (2011).}
\label{eq_pinned}
\end{figure}

Consider a particle in suspension near a planar fluid interface. When the particle attaches to the interface, it eliminates a patch of solid-liquid contact ${\Delta A_{SL}}$ and makes a hole in the interface of area ${\Delta A_{LV}}$.  Furthermore, it can make distortions in the surrounding interface with area  ${\delta A}$. The net energy change or trapping energy is:
\begin{eqnarray}
\Delta E =  ({\gamma _{SL}-\gamma _{SV}}){\Delta A_{SL}}+{\gamma}{\Delta A_{LV}}+{\gamma }{\delta A}.
\end{eqnarray}
where  $\gamma _{SL}$ and  $\gamma _{SV}$ are the surface energies of the solid-liquid and solid-vapor surfaces.  When the trapping energy is large compared to $k_B T$, the particle is trapped, i.e., it cannot spontaneously leave the interface. We discuss the trapping energy for two scenarios shown in Figure \ref{eq_pinned}.
\subsection{Trapping energy for a sphere at equilibrium} 
The case of a perfect sphere at equilibrium  with contact angle $\theta _0$ is an important ideal limit. The particle can attach without deforming the surrounding interface, so $\delta A=0$, and the contact line is simply a circle in the plane of the interface \cite{Pieranski}. The trapping energy is:
\begin{eqnarray}
\Delta E  =  - {\gamma}\pi {a^2}{(1 - \left| {\cos {\theta _0}} \right|)^2}.
\end{eqnarray}
where  $\theta _0$ is defined by the balance of the surface energies tensions given by the Young's equation,
\begin{eqnarray}
\cos \theta_0=\frac{\gamma_{SV}-\gamma_{SL}}{\gamma}.
\end{eqnarray}
 By attaching, the particle reduces the area of the liquid vapor interface, lowering the system energy. This effect is modulated by the particle wetting properties. The trapping energy is remarkably large. For example, for air-water interfaces, the surface tension is $\gamma=72~{{{{mN}}} \mathord{\left/{\vphantom {{{{mN}}} {{m}}}} \right.\kern-\nulldelimiterspace} {{m}}}$ or $18~{{{{{k}}_{{B}}}{{T}}} \mathord{\left/
 {\vphantom {{{{{k}}_{{B}}}{\rm{T}}} {{{n}}{{{m}}^{{2}}}}}} \right.
 \kern-\nulldelimiterspace} {{{n}}{{{m}}^{{2}}}}}$. Typical trapping energies for microparticles can be $10^5-10^6~k_BT$.
\subsection{Contact Line Pinning}
This simple picture is complicated by contact line pinning on nanoscopic sites of roughness or chemical heterogeneity, for which there is now strong experimental evidence, even for simple, apparently homogeneous spherical polystyrene microparticles \cite{Kaz:2012pin}. Pinned contact lines alter trapping energies fundamentally, as particles with undulated contact lines distort the interface around them, so ${\delta A} \neq 0$.  The implications of contact line distortion on isolated trapped particles are still being resolved \cite{Kaz:2012pin, Boniello, WangSM2016}. For isolated particles, contact line pinning changes the trapping energy by the amount $\gamma {\delta A}$; this is termed a self-energy contribution \cite{Kralchevsky2000145}. As we discuss below, particle-sourced interface distortions are the source of capillary pair interactions.  Here, we review recent literature on contact line pinning for particles at interfaces.\\ 
\indent Spherical microparticle attachment  to the interface occurs via a rapid snap-in event, followed by a slow wetting as the contact line moves toward equilibrium \cite{Kaz:2012pin, WangSM2016, CheAPL2012, ColosquiPRL2013, SinghPNAS}.  Snap-in includes a rapid opening or breaching of the interface, and as well as the formation of a contact line on the particle surface. Immediately after snap-in, particles oscillate owing to inertia associated with this event. In experiments with glass microbeads with various surface chemistries and radii, snap-in occurred within 0.1ms, reflecting a balance of inertia and surface tension independent of particle wetting; the snap-in force however, depended on particle size and wetting \cite{CheAPL2012}. Once the oscillations end, the contact line exhibits slow, glassy dynamics as it approaches equilibrium \cite{Kaz:2012pin, WangSM2016}. Contact line motion occurs with negligible capillary number $Ca$ based on contact line velocity, so viscous effects, typically important for contact line motion in dynamic spreading \cite{cox} are negligible. Using holographic imaging to track the position of a polystyrene colloid near a decane-water interface, the contact line was inferred to move slowly, i.e., logarithmically in time for particles close to equilibrium.  Because of these slow kinetics, similar particles with differing breach times have different wetting positions in the interface.  The observed logarithmic relaxation is consistent with a model in which the contact line is pinned at nanoscopic heterogeneities with hopping frequencies given in terms of the Blake-Haynes model from molecular kinetic theory \cite{Blake1969}. Further analysis and supporting molecular dynamics simulation reveals contact line relaxations are initially exponential with a visco-capillary time scale, then exhibit slow logarithmic aging like that captured in experiments, and finally exhibit exponential decay to equilibrium, suggesting that the glassy dynamics end \cite{ColosquiPRL2013}. Recent studies of particles made of different materials show that the glassy contact line dynamics are generic, but the energy of pinning sites varies strongly for different materials as presented in Figure \ref{eq_pinned}c \cite{WangSM2016}. \\
\indent Contact line pinning may have important dynamical consequences even for isolated particles attached at the interface; pinned states, and associated enhanced dissipation owing either to contact line hopping or interaction of the undulated interface with capillary waves were invoked in a recent study reporting unexpectedly high drag on Brownian particles adsorbed from suspension in air at air-water interfaces \cite{Boniello}.
\subsection{Trapping energy for a particle with a pinned contact line}
\begin{figure}[h]
\includegraphics[scale=0.48]{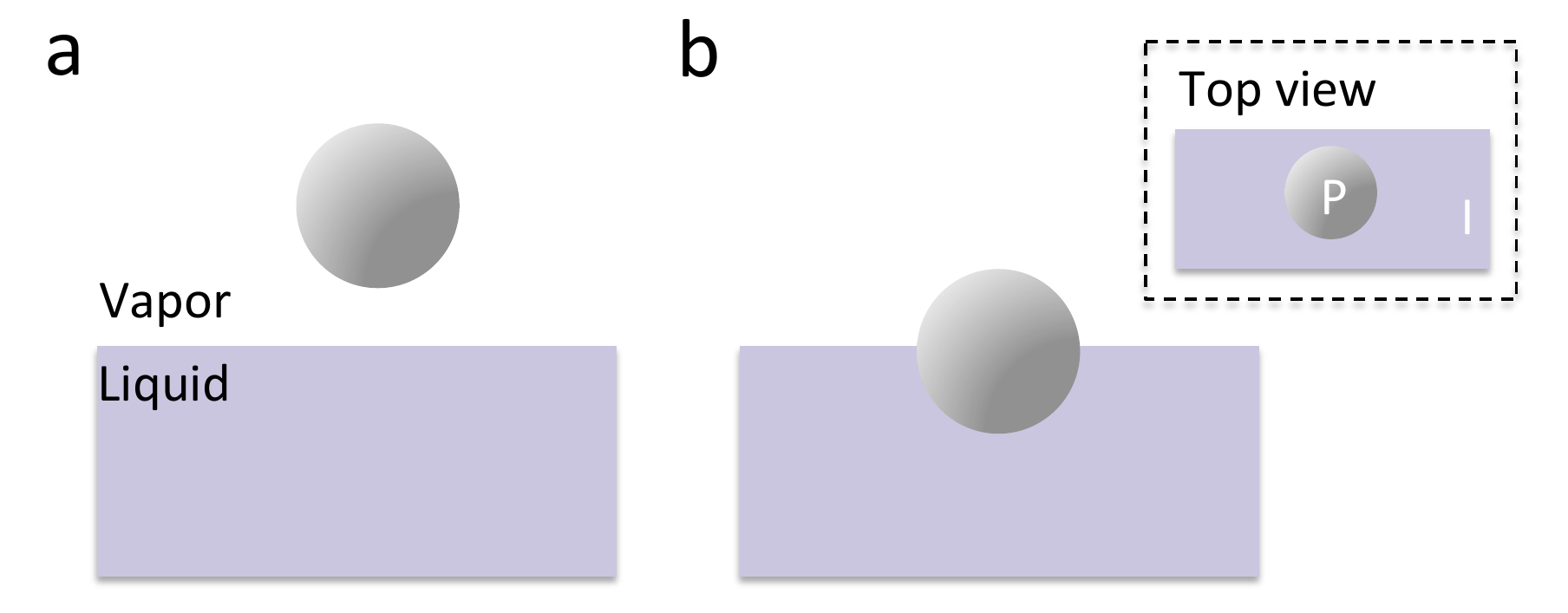}
\caption{Schematic of a spherical particle (a) immersed in vapor phase and becomes (b) adsorbed to a liquid-vapor interface. Inset: top view of a particle (P, particle domain) on an interface (I). }
\label{attach}
\end{figure}

For spherical particles with pinned contact lines, there are several unknown aspects that complicate evaluation of the trapping energy.  These include the unknown angle characterizing the degree of immersion of the particle in the fluid, $\theta _{tr}$, the unknown contact line shape, and the associated area of particle-sourced deformation in the interface $\delta A$. In the limit of small slopes, the shape of the contact line can be decomposed into Fourier modes \cite{Stamou}, and $\delta A$ can be found by determining the height $h$ of the interface around the particle above the reference plane and evaluating the associated area for each mode. The shape of the interface is given by a decaying multipole expansion expressed in a polar coordinate system $(r,\phi)$ in the plane of the interface with origin at the center of the hole made by the particle in the interface. Each order of this expansion is excited by the corresponding Fourier mode at the contact line: 
\begin{eqnarray}
h(r,\phi )={b_0}\ln r + \sum\limits_{m = 1}^\infty  { {c_m}{r^{ - m}}\cos (m\phi+\alpha_m) }.
\end{eqnarray}
where $\alpha_m$ is the phase angle of mode $m$. Since body forces and torques on the particles are negligible, mechanical equilibrium requires that ${b_0}={c_1}=0$. Thus the quadrupolar distortion is the first surviving mode in the interface distortion \cite{Stamou}. Letting $h_{qp}$ be the amplitude of this mode, the interface height to leading order is:
\begin{eqnarray}
h(r,\phi ) = {h_{qp}}\frac{{{a^2}}}{{{r^2}}}\cos 2\phi.
\end{eqnarray}
This term describes the long-range interface distortion from any particle with any undulated contact line; the existence of this mode gives rise to universal behavior between interacting particles in the far field, and for small particles at curved fluid interfaces.\\
\indent To evaluate the trapping energy for a particle, $\delta A$ must be evaluated. Dividing the particle-free interface into two domains, $I$ and $P$, where the domain  $P$ is occupied by the particle after attachment, and the domain  $I$ is outside of the contact line (Fig. \ref{attach}), the area can be evaluated in  the limit of small slopes:
\begin{eqnarray}
\delta {A} \approx \mathop{{\int\!\!\!\!\!\int}\mkern-21mu }\limits_I {  \;\;\; \frac{\nabla {h} \cdot \nabla {h}}{2}r{\rm{d}}r{\rm{d}}\phi}=\pi h_{qp}^2.
\end{eqnarray}
The trapping energy for the particle can be evaluated thus:  
\begin{eqnarray}
{\Delta E_{planar} = - {\gamma}\pi {a^2}(1 - \left| {\cos {\theta _{tr}}} \right|)^{2} + {\gamma}\pi h_{qp}^2}\label{p1}.
\end{eqnarray}
The first term is similar to the equilibrium case except that the angle characterizing the degree of immersion in the trapped state $\theta_{tr}$ replaces $\theta_{0}$. The second term is the 'self energy' of the particle, i.e. the energy cost associated with the area of the distortion around the particle. Similar terms appear from the higher order modes in the multipole expansion \cite{Stamou, Danov_bipolar}. 
\subsection{Nonspherical Particles}
The contributions to the trapping energy remain the same for complex shaped particles. These include the energy decrease caused by the hole in the interface, modulated by the particle wetting energies, and the energy owing to the excess area of the interface distortion. The evaluation of each contribution is difficult for many reasons. The equilibrium wetting configuration of complex shaped particles cannot typically be derived analytically \cite{Lewandowski2010,Botto2012Review,Botto2012SM}. Thus, the size of the hole in the interface, the height of the particle, and the interface distortion must typically be found via simulation. Furthermore, contact line pinning occurs on these particles, so, though simulated equilibrium wetting configurations lend guidance, they do not suffice to predict the dynamic state of the interface and the particle. There are, however, important simplifications that can be made to make some progress.  Pinned contact lines around non-spherical particles can be decomposed into Fourier modes; interface distortions can be described in multipole expansions.  For elongated particles, an expansion in ellipsoidal coordinates for a particular particle aspect ratio is appropriate \cite{Loudet2011}.  The quadrupolar modes in this coordinate accurately describes the far field interface shape for elongated particles \cite{Lewandowski2010}. 
\subsection{Key Findings for Isolated Particles} 
\begin{figure*}
\includegraphics[scale=0.5]{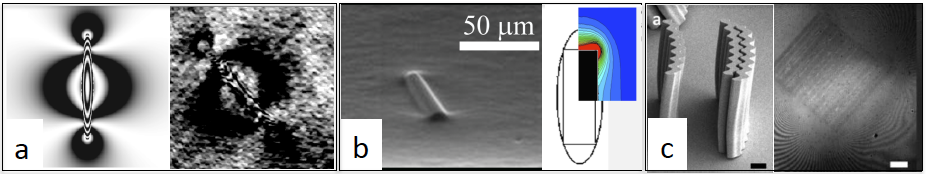}
\caption{Nonspherical particle deformation on planar interfaces. (a) Simulated profile of an isolated ellipsoid (Aspect Ratio or AR=6.9) comparing to the Phase Shifting Interfereometry (PSI) image. Reprinted with permission from Reference \cite{Loudet2006PRL} (https://doi.org/10.1103/PhysRevLett.97.018304); copyright 2006 by the American Physical Society. (b) Environmental SEM image of a SU-8 microcylinder (AR=3) on a gelled fluid interface and simulated excess area map of a cylindrical particle. Adapted with permission from \cite{Lewandowski2010}. Copyright 2010, ACS. (c) SEM image of wavy particles and their deformation at the air-water interface; scale bar 50 $\mu m$. Adapted from Reference \cite{Lu2013SM} with permission from The Royal Society of Chemistry.}
\label{ns}
\end{figure*}

Anisotropically shaped microparticles attach to interfaces via a process related to that for spheres. However, the highly non-uniform dynamic contact angle along the contact line contour as the particle enters the interface gives interesting dynamics \cite{manoharanpreprint}. An ellipsoidal microparticle enters the interface via a rolling motion consistent with non-uniform displacement of contact line segments; segments with greater differences between the $\theta_{tr}$ and $\theta_{0}$ moved faster. These observations imply that differing, time dependent surface deformations can be made by similar particles. Complex shaped particles, whether they are adsorbed from suspension or spread via solvents can assume a variety of configurations \cite{Lewandowski2010, Loudet, Loudet2006PRL, Lewand2009SM, SharpRSC2014, Vermant}. As a crude guiding principle, isolated particles often assume orientations in which they make the largest hole in the interface; e.g. elongated particles have their long axes in the plane of the interface. Examples shown in Figure \ref{ns}a-b include prolate ellipsoids spread using isopropyl alcohol at water-air interfaces \cite{Loudet, Loudet2006PRL, Vermant}, right circular cylinders placed at aqueous-air interfaces via spreading solvent \cite{Lewandowski2010, Lewand2009SM, Lewand2008Langmuir} or adsorbed from suspension at hexadecane-water interfaces. In contrast, thin disks typically adsorb at aqueous-oil interfaces with their circular face in the interface \cite{Yaodisk}.\\
\indent  Anisotropic microparticles can make very strong distortions in fluid interfaces that can be imaged via interferometry \cite{Loudet2006PRL, Lewandowski2010} and compared to simulated equilibrium wetting configurations \cite{Loudet}. Distortions around prolate ellipsoids and cylinders have quadrupolar symmetry. Observed interface shapes agree well with the height of  quadrupolar modes in elliptical coordinates within a few radii of contact with the particle \cite{Lewandowski2010, Botto2012SM}, a fact that facilitates analysis. In the very near field, however, only simulations capture details \cite{Lehle}, in particular near features like sharp edges and corners. In most simulations \cite{Botto2012SM, SolignoPRL2016}, equilibrium contact angles are assumed in the near field, in spite of the importance of pinned contact lines. Finer features also play a role, including particle roughness \cite{Lucassen1992}. To investigate these effects, particles designed to form wavy contact lines with wavelength and amplitude small in comparison with the particle length have been studied \cite{Lu2013SM}; the distortions made by the wavy features decay over distances similar to their wavelength and change the energy landscape only in the near field around the particle (Fig. \ref{ns}c). These will have implications in pair interactions.
\\
\indent Finally, while the quadrupole is the leading order distortion for the interface in the far field absent external forces, there are particle configurations that  excite higher order modes, e.g. simulations of cuboid-shaped particles show that, in certain orientations, particles excite well-defined hexapolar modes in the interface \cite{SolignoPRL2016}. 
\subsection{Summary} 
Particles become trapped at planar fluid interfaces. Perfectly smooth spheres with equilibrium wetting conditions leave the interface around them unperturbed. However, particles with pinned contact lines, patchy wetting or non-spherical shapes distort the interface around them. The associated self-energies contribute to the trapping energy of the particle. Distortions due to various particle features decay at different distances from the particle.  All particles make quadrupolar distortions in  the far field; higher order modes owing to complex contact line shapes decay more rapidly.  Within a few particle radii of contact, features like particle elongation become apparent; closer still, contact line waviness due to particle geometry, roughness or pinning, and sharp edges play a role \cite{Lu2013SM}. The distortion fields around microparticles at interfaces play a central role in pair interactions, as is discussed in section \ref{Secprint} below.
\section{Pairs of particles on planar interfaces} \label{Secprint}
Particles interact at fluid interfaces to minimize the interfacial area. Interactions between capillary  multipoles are often likened to those between charge multipoles except that like charges attract. In this rubric, regions above the reference plane are positive while those below the reference plane are negative. When distortions $h$ from neighboring particles overlap, particles orient and migrate so that regions with "like charge" overlap. In this way, the slope of the interface, and the area $\delta A$ owing to the distortions, decreases.
(We revisit the analogy to electrostatics in section 5.1, below.) As particles approach, different parts of their distortion fields interact depending on their separation distance. Interactions first occur because of their quadrupolar modes.  Closer to contact, higher order modes, near field distortions, and the presence of corners and edges play roles. Furthermore, rearrangement of wetting configurations and contributions from the associated solid-liquid wetting energies can, in principle, play a role. All of these have been discussed in prior literature \cite{Stamou, Kralchevsky2000145, Botto2012SM, Danov_bipolar, Botto2012Review}. Therefore, we discuss main concepts and findings in only a cursory manner here, and refer the interested reader to those reviews for a more thorough treatment.\\
\indent Below, we derive the capillary energy of interaction between two colloidal particles with pinned contact lines on an otherwise planar interface, using the method of reflections, and compare  this to the exact solution in bipolar coordinates.  This treatment differs from the seminal work in the literature \cite{Stamou} in that we do not adopt the superposition approximation. 
\subsection{Method of Reflections}
\begin{figure}[h]
\includegraphics[scale=0.26]{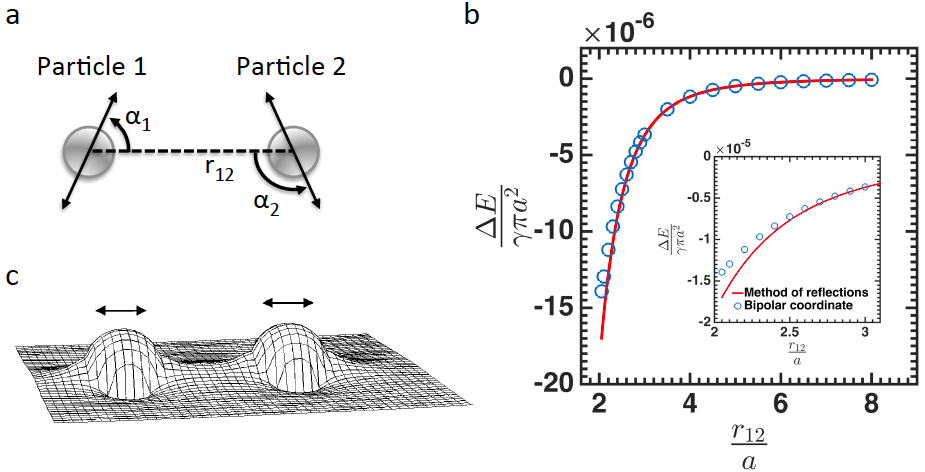}
\caption{Schematic of pair of particles interacting with their quadrupoles. (a) Spheres separated by distance $r_{12}$ align in mirror symmetric orientation where the doublesided arrows indicate the rise of the interface, and  $\alpha_1$ and $\alpha_2$  define the phase angle for particle 1 and 2, respectively. (b) Non-dimensional interaction energy comparison between the method of reflections and the exact solution from bipolar coordinate calculation. Inset: Near field comparison of the two methods \cite{Varenna}. (c) A pair of interacting spheres with their quadrupoles co-aligned (rise-to-rise). Reprinted with permission from Reference \cite{Stamou} (https://doi.org/10.1103/PhysRevE.62.5263); copyright 2000 by the American Physical Society. }
\label{pair_int}
\end{figure}

Particles $1$ and $2$ of radius $a$ are separated by distance $r_{12}$, with ${a \mathord{\left/
 {\vphantom {a {{r_{12}}}}} \right.
 \kern-\nulldelimiterspace} {{r_{12}}}}\ll 1$ (see Fig. \ref{pair_int}a). Both particles have pinned contact lines with quadrupolar modes of amplitude $h_{qp}$. In isolation, the shape of the interface around each colloid can be expressed in terms of polar coordinates $(r_1, \phi_1)$ and $(r_2, \phi_2)$ located at the centers of the particles:
\begin{align}
&h_1 =  h_{qp}\frac{{a^2}}{{r_1^2}}\cos 2({\phi _1}- {\alpha}_1)
\end{align}
and
\begin{align}
&h_2 = h_{qp} \frac{{{a^2}}}{{r_2^2}}\cos 2({\phi _2}-\alpha_2),
\end{align}
where $\alpha_1$ and $\alpha_2$ denote the phase angles of particles with respect to the line connecting particle centers. A Taylor Series expansion of the distortion owing to  particle $2$ near  particle $1$ yields:
 \begin{align}
 {h_2} = {\left. {{h_2}} \right|_{{{\bf{r}}_{12}}}} + {\bf{r}}_1 \cdot {\left. {\nabla {h_2}} \right|_{{{\bf{r}}_{12}}}} + {\bf{r}}_1 \cdot {\left. {\frac{{\nabla \nabla {h_2}}}{2}} \right|_{{{\bf{r}}_{12}}}} \cdot {\bf{r}}_1 + ...,\label{taylor}
 \end{align}
where ${\bf{r}}_1$ is the position vector from the origin at particle $1$ and ${\bf{r}}_{12}$ is the vector from the origin to particle $2$. In the above expression, the first two terms are changes to the height and slope; absent body forces and torques, particle $1$ adjusts its height and tilt to eliminate these terms. The third term is the curvature field created by particle $2$ in the vicinity of particle $1$.  This term is the leading order distortion made by particle $2$ near $1$, and defines the far field distortion ${h_{\infty}}$ for particle 1 \cite{KimKarrila, HappelBrenner}. The shape of the interface around particle $1$ in the plane tangent to the interface can be found by solving this boundary value problem: 
 \begin{align}
&{\nabla ^2}{h_1} = 0,
\end{align}
with the boundary condition at the contact line:
 \begin{align}
&{h_1}({r_1} = a) = {h_{{{qp}}}}\cos 2({\phi _1} - {\alpha _1}),
\end{align}
and in the far field,
  \begin{align}
 {h_1(r_1 \to \infty )}=h_{\infty}=3h_{qp} \frac{a^2}{{{r_{12} ^4}}}r_1^2\cos 2({\phi _1}+\alpha_2).
\label{curv2}
 \end{align}
Solving for the interface shape around the particle, 
 \begin{align}
&{h_1} = 3h_{qp} \frac{{a^2}}{{r_{12}^4}}r_1^2\cos 2({\phi _1} + {\alpha _2}) + {\eta _1}
\end{align}
and
 \begin{align}
&{\eta _1} = {h_{{{qp}}}}\frac{{{a^2}}}{{r_1^2}}\cos 2({\phi _1} - {\alpha _1}) - 3{h_{{{qp}}}} \frac{{{a^2}}}{{r_{12}^4}}\frac{{{a^4}}}{{r_1^2}}\cos 2({\phi _1} + {\alpha _2}).
\end{align}
The disturbance ${\eta _1}$ includes the particle sourced term and an induced or reflected term ``undoing'' the curvature created by particle $2$. \\ 
To calculate the area associated with this disturbance field $\delta {A_{1}}$ around particle 1, we evaluate:
\begin{align}
\delta {A_{1}} \approx \mathop{{\int\!\!\!\!\!\int}\mkern-21mu }\limits_I {  \;\;\; \frac{\nabla {h_1} \cdot \nabla {h_1}}{2}r_1{\rm{d}}r_1{\rm{d}}\phi_1}-\mathop{{\int\!\!\!\!\!\int}\mkern-21mu }\limits_I {  \;\;\; \frac{\nabla {h_{ \infty}} \cdot \nabla {h_{ \infty}}}{2}r_1{\rm{d}}r_1{\rm{d}}\phi_1}.
\end{align}
$\delta {A_{1}}$ contains three terms; the self energy that occurs for isolated particles, a pair interaction energy that depends on $r_{12}$, and a higher order term that makes negligible contributions to leading order. Taking the difference in energies for finite $r_{12}$ and for infinite separations, the capillary energy for particle 1 interacting with particle 2 to leading order is: 
\begin{align}
\Delta {E_1}=  -6 \gamma \pi {h_{{{qp}}}}^2 \frac{{{a^4}}}{{r_{12}^4}}\cos 2({\alpha _1} + {\alpha _2}).\label{pair-22}
\end{align}
Particle $2$ has an identical contribution, so:
\begin{align}
\Delta E = 2\Delta {E_1}=-12 \gamma \pi {h_{{{qp}}}}^2 \frac{{{a^4}}}{{r_{12}^4}}\cos 2({\alpha _1} + {\alpha _2}).\label{pairint} \end{align} 
The interface shape can be solved exactly in bipolar coordinates \cite{Danov_bipolar, Arfken} for pinned quadrupolar contact lines on both particles. The result, however, includes changes in height and slope from the neighboring particle that are not present in the absence of body forces and torques. By amending the boundary condition at the contact line to remove these effects, the interface height and the excess area around both particles 1 and 2, $\delta A$ can be calculated analytically.  The resulting interaction energy is compared to the pair interaction energy in Equation \ref{pairint} in Figure \ref{pair_int}b. The two solutions agree for particles more than a radius from contact. However, very near to contact, there is a deviation between the two solutions.  This deviation indicates the importance of higher-order reflections in the near field.
\\   
{\indent}The pair-interaction energy between particles predicts that particles attract only if they are in mirror-symmetric orientations i.e., $\alpha_1+\alpha_2=0^{\circ}$ (Fig. \ref{pair_int}c). Particles that are mis-aligned rotate to assume mirror-symmetry, and then migrate. 
For common fluid interfaces, $\gamma \sim 10 k_BT/ nm^2$.  Particles with contact line distortions as small as $2nm$ can have more than $10k_BT$ of interaction at center to center separations of several radii.  Microparticles with rough surfaces, anisotropic shapes or patchy wetting can have contact lines with far larger amplitude modes. The attractive capillary force corresponding to the energy expression of Equation \ref{pairint} $\sim r_{12}^{-5}$. For particles moving in creeping flow, this force is counterbalanced by viscous drag $\sim dr_{12}/dt$.  This balance requires that particles move with a power law $r_{12} \sim (t_f-t)^{1/6}$ \cite{Loudet}.  
\subsection{Key Findings of Interactions on Planar Interfaces} 
\begin{figure*}
\includegraphics[scale=0.45]{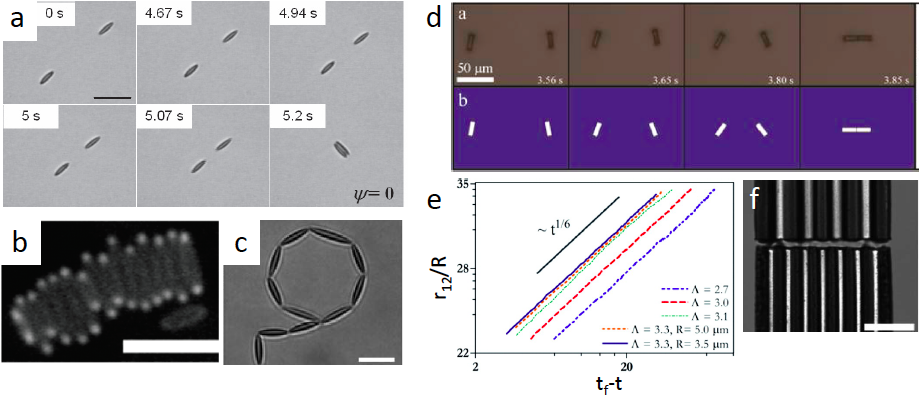}
\caption{Interactions of complex shaped particles. (a) Time lapsed optical microscopy image of a pair of interacting ellipsoids trapped at the air-water interface. Scale bar: 19.4 $\mu m$. Reproduced with permission from Reference \cite{LoudetArrows}. Copyright IOP Publishing \& Deutsche Physikalische Gesellschaft. CC BY-NC-SA (b) Smectic-like assembly of sterically stabilized PMMA ellipsoids in side-to-side conformation. Scale bar: 10 $\mu m$ Reprint with permission from \cite{Vermant}. Copyright 2010, ACS. (c) Tip-to-tip assembly of polystyrene ellipsoids; scale bar 13.6 $\mu m$. Reprinted with permission from Reference \cite{Loudet} (https://doi.org/10.1103/PhysRevLett.94.018301); copyright 2005 by the American Physical Society. (d) Time lapsed image of cylindrical SU-8 micro cylinders assemble in mirror symmetric orientation to form an end-to-end chain (top: optical microscopy image; bottom: simulation).  (e) Power law dependence of separation distance as a function of contact time, where $t_f$ is the time in seconds when the pair of cylindrical particles contact end to end. (d)-(e) Adapted with permission from \cite{Lewandowski2010}. Copyright 2010, ACS.  (f) Wavy SU-8 particles of mismatched wavelengths assembled into a structure with an equilibrium separation distance. Scale bar: 100 $\mu m$. Adapted from \cite{Lu2013SM} with permission from The Royal Society of Chemistry.}
\label{pair_ell}
\end{figure*}

The strong deformation fields around anisotropic particles make them excellent vehicles for studying capillary interactions. Here we focus on ellipsoids and cylinders, and the role of particle roughness. The dynamics of microparticle assembly were first observed for ellipsoids at a water-oil interface a decade ago \cite{Loudet}, motivating interest in ellipsoidal particle assembly \cite{DasguptaLang2014, Lehle, LoudetArrows, Botto2012SM, Loudet2011}. The particles, with major axis $\sim10 \mu m$ and minor axes ~$\sim2 \mu m$ interacted over distances as great as six particle lengths with weakly Brownian trajectories in the far field, and well determined paths in the near field. Particles approached in either tip-to-tip and side-to-side configurations (Fig. \ref{pair_ell}a-c). For tip-to-tip interactions, the particles obeyed the expected power law for interacting polar quadrupoles in the far field. For side-to-side arrangements, a lower exponent was reported; subsequent detailed simulation shows that contributions from higher order modes are significant at separations as large as four particle lengths \cite{Lehle}. The capillary energy change along a trajectory, inferred from viscous dissipation, was very strong, $\sim10^4 k_BT$. Particles assemble tip-to-tip for polystyrene particles or side-to-side for silica coated particles \cite{Loudet}. The existence of these preferred alignments is an interesting feature, as the theory described in section \ref{Secprint} does not predict this effect. Theory for pairs of interacting elliptical quadrupoles does predict mirror symmetric approaches, tip-to-tip assembly at contact, and rotation to side-to-side alignment after contact \cite{Lewandowski2010}.  Detailed simulation of the interface between ellipsoids near contact also predicts side-to-side assembly \cite{Lehle,Botto2012SM}. Thus, near field capillary interactions favor side-to-side arrangements. The tip-to-tip arrangement might be enforced by electrostatic repulsion, known to be significant for polystyrene particles at water-oil interfaces. Indeed, particles with scant surface charge (PMMA or poly(methyl methacrylate) microparticles) assemble side-to-side, as shown in Figure \ref{pair_ell}b \cite{Vermant}. Similar arrangements have been reported for diverse ellipsoidal shaped objects, including mosquito eggs \cite{Loudet2011} and whirligig beetles \cite{Voise1357}. The mechanics of these structures is rich. Particles chained tip-to-tip rotate freely while maintaining contact, while chains of side-to-side particles can bend weakly under compression. 

\indent Crowded interfaces of ellipsoids also have interesting behavior that we describe only briefly. Simulation of weakly non-spherical particles assuming pairwise additivity suggests that even nanometric deviation from sphericity can drive capillary assembly of ellipsoids into a variety of structures that include dendritic-trapped configurations, rafts and hexagonal lattices \cite{NieropEPL2005}. The rheology of rafts of ellipsoidal particles in tip-to-tip assembly differs from sphere-laden interfaces; particle monolayers are elastic at low surface area fractions and yield via a series of flipping events under compression \cite{MadivalaLang2009}, and this has major implications in important processes like convective assembly within evaporating drops \cite{YunkerPRL}. \\
\indent Cylindrical microparticles interact over distances comparable to ten particle lengths, with excellent agreement with the power law for interacting polar quadrupoles in the far field as presented in Figure \ref{pair_ell}d-e \cite{Lewandowski2010}. If the particles are already oriented end-to-end, they maintain that alignment until contact. However, if they are oriented side-to-side, they rotate while maintaining mirror symmetry to assemble and form a straight and rigid dimer. On sparse interfaces, long chains comprising many cylinders form.\\  
\indent The chains of cylinders are remarkably rigid, failing to bend or break even under significant torque applied by rotating them in a magnetic field. To understand this effect, cylindrical microparticles near contact were compared to ellipsoidal particles near contact in simulation \cite{Botto2012SM}. Cylinders were simulated in an end-to-end alignment, and then in arrangements where they rotated from that end-to-end alignment (Fig. \ref{pair_ell}d). The steric barrier posed by the particle's sharp edges and the associated rearrangements of the capillary bridge between the particles give rise to a strong energy barrier to rotation that enforces the rigid end-to-end alignment. These features are absent on ellipsoids, which can roll over each other freely near contact; the associated energy landscape indicates that the capillary bond between pairs of ellipsoids is elastic, so chains of side-to-side ellipsoids are flexible, whereas chains of cylinders should remain aligned until they snap under applied torque.  \\  
\indent It is interesting to ask whether particles come to contact, and what would limit their proximity. In an early study, particle roughness was suggested as a source of repulsive capillary interactions.  Wavy contact lines pinned on the rough sites would create local disturbances near the particle, which is important only in the very near field \cite{Lucassen1992}. When neighboring particles approach, these disturbances would interact. If they matched perfectly, with identical wavelengths, phases and amplitudes, particles would attract.  However, if they differ, as would be expected for random roughness, particles would be repelled. This concept was recently demonstrated using particles with wavy edges \cite{Lu2013SM}. In the far field, these particles experience the usual capillary attraction. However, when distortions from the wavy contact lines overlap, particles with differing undulations are repelled, as shown in Figure \ref{pair_ell}f.

\subsection{Summary}
Particles with pinned contact lines interact via capillarity over remarkable distances.  Anisotropic particles align as they migrate to preferred configurations. In the near field, details in the particle shape play major roles in determining the strength of the interactions, and the distance of closest approach.  In his analysis of particle interactions in the far field, Stamou noted that one particle moved in the curvature field of its neighbor \cite{Stamou}. In principle, however, any means of pinning or distorting  the interface far from the particle can create a curvature field. It is a natural extension to consider particles on curved interfaces.
\section{Curvature capillary energy}\label{curved-interface}
\begin{figure}[h]
\includegraphics[scale=0.7]{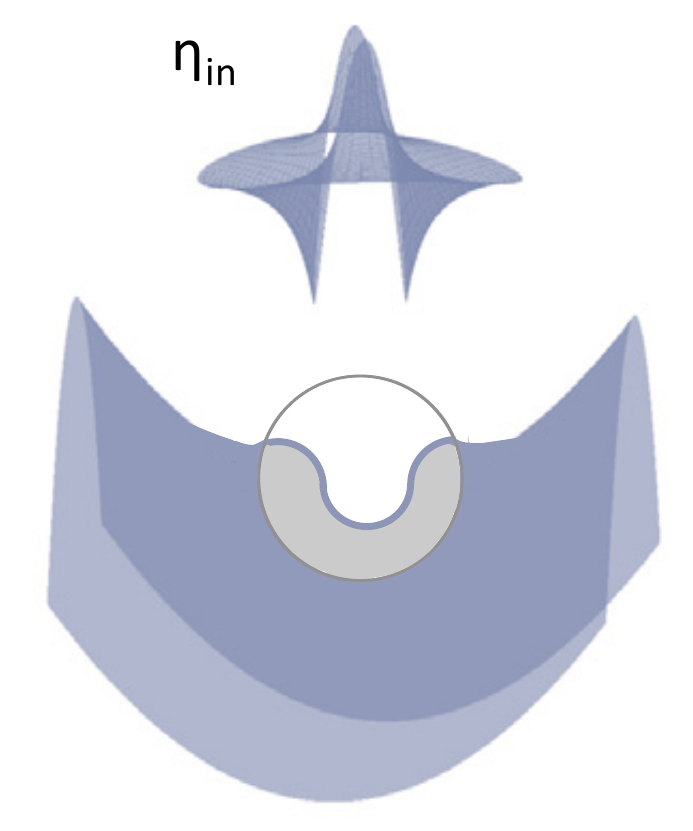}
\caption{Schematic of a spherical particle with quadrupolar deformation on a curved interface where the particle-induced distortion is denoted $\eta_{in}$.}
\label{induced}
\end{figure}

When a particle with a pinned, undulated contact line attaches to a curved fluid interface, the interface curvature alters the trapping energy. In the limit of small slopes, for particles small compared to the principal radii $R_1$, $R_2$, the host interface near the particle center can be expanded in terms of the mean curvature $H_0=\frac{1}{2}(c_1+c_2)=\frac{1}{2}(\frac{1}{R_1}+\frac{1}{R_2})$ and the deviatoric curvature $\Delta c_0= c_1-c_2=\frac{1}{R_1}-\frac{1}{R_2}$, where $c_1$ and $c_2$ are the principal curvatures evaluated at the particle center of mass. In  the absence of the particle, the interface shape is
\begin{align}
{h_{0}} = \frac{{{{r^2}H_0}}}{2} + \frac{{\Delta {c_0}}}{4}{r^2}\cos 2\phi.
\end{align}
When a particle attaches to the curved interface, the trapping energy is: 
\begin{align}
\Delta E = (\gamma _{SL}-\gamma _{SV})\Delta A_{SL}  + \gamma \Delta A_{LV}  +\gamma \delta A  + ~{\rm{  PV~work}}.
\end{align}
We consider the right-hand side of this expression term by term. The wetting energies are unchanged from the planar case given the symmetries of the Fourier modes that describe the contact line. The hole made by the particle in the interface and the area in the distortion field both depend on the curvature field, and must be computed. Finally, changes in height owing to the particle require ${PV}$ work against the pressure jump at the interface. To evaluate these terms, we find the disturbance made to the interface shape by the particle $\eta=h-h_{0}$, which requires solution of a simple boundary value problem:
\begin{align}
{\nabla ^2}h & = 0,\\
h(r = a) & = {h_{qp}}\cos 2\phi ,\\
h(r \to \infty ) & = {h_{0}},\\
h & = {h_{0}} + \eta_{qp}+\eta_{in}\\
&= {h_{0}}+ h_{qp} \frac{{{a^2}}}{{{r^2}}}\cos 2\phi- \frac{{{a^2}\Delta {c_0}}}{4}\frac{{{a^2}}}{{{r^2}}}\cos 2\phi +\omega_0. \nonumber
\end{align}
The disturbance $\eta$ has two parts: the particle imposed distortion $\eta_{qp}$ and the induced disturbance or reflected mode ${\eta _{in}}$ (Fig. \ref{induced}). Additionally, the particle shifts vertically to situate itself in the interface with finite  mean curvature $\omega_0=\frac{a^2H_0}{2}$; this requires PV work:  
\begin{align}
\Delta P\mathop{{\int\!\!\!\!\!\int}}\limits_P 
 {({h_{0}} - \frac{{{H_0}{a^2}}}{2})r dr d\phi}  =  - {\gamma\pi}{a^2}\frac{ H_0^2{a^2}}{2}.
\end{align}
The area of the interface is given by the sum $\Delta {A_{LV}}+\delta A$. By attaching to the interface, the particle forms a circular hole  with area $\pi a^2$, with a correction owing to curvature:
 \begin{align}
 &\Delta {A_{LV}}=- \mathop{{\int\!\!\!\!\!\int}}\limits_P 
 {(\frac{{\nabla {h_{0}} \cdot \nabla {h_{0}}}}{2})r dr d\phi}  =  - \pi {a^2}(1+\frac{{{a^2}H_0^2}}{4} + \frac{{{a^2}\Delta c_0^2}}{{16}}).\label{host-11}
 \end{align}
To evaluate $\delta A$, several contributions must be evaluated.
 \begin{align}
 \delta A  = & \mathop{{\int\!\!\!\int}}\limits_I 
 {(\frac{{\nabla {\eta _{in}} \cdot \nabla {\eta _{in}}}}{2})r dr d\phi} + \mathop{{\int\!\!\!\int}}\limits_I  {({{\nabla {\eta _{in}} \cdot \nabla {\eta _{qp}}}})r dr d\phi} \nonumber \\
& + \mathop{{\int\!\!\!\int}}\limits_I  {(\frac{{\nabla {\eta _{qp}} \cdot \nabla {\eta _{qp}}}}{2})r dr d\phi}+ \mathop{{\int\!\!\!\int}}\limits_I 
 {({{\nabla {\eta} \cdot \nabla {h_{0}}}})r dr d\phi}.\label{int-1K}
\end{align}
The first term in this expression is the area from the induced disturbance around the particle. The divergence theorem requires that this term be equal and opposite to the deviatoric curvature correction to $\Delta A$.  The second term  is the interaction of two disturbance terms: 
\begin{align}
\mathop{{\int\!\!\!\!\!\int}}\limits_I
 {({\nabla {\eta _{qp}} \cdot \nabla {\eta _{in}}})r dr d\phi}  =  - \frac{\pi }{2}\Delta {c_0}{a^2}{h_{qp}}.
\end{align}
The third term is the area from the particle sourced disturbance, evaluated previously, and the final term is identically zero. Gathering terms, the trapping energy on a curved  interface is: 
\begin{align}
\Delta E = \Delta {E_{planar}} - {\gamma }\pi {a^2}(\frac{{3{a^2}H_0^2}}{4} + \frac{{{h_{qp}}\Delta {c_0}}}{2}),\label{energy-221}
\end{align}
where $\Delta {E_{planar}}$ is defined in  Eq.~\ref{p1}. The trapping energy is reduced by interface curvature.  In this discussion, we assumed that the particle's quadrupolar mode was aligned with the saddle shape of the interface.  If this were not the case, there would be a capillary torque exerted on the particle which would cause the particle to align \cite{Lewand2008Langmuir}, and the energy expression becomes: 
\begin{align}
\Delta E = \Delta {E_{planar}} - {\gamma }\pi {a^2}(\frac{{3{a^2}H_0^2}}{4} + \frac{{{h_{qp}}\Delta {c_0}}}{2}\cos 2 \alpha),\label{energy-221}
\end{align}
where $\alpha$ is the angle between the quadrupolar rise axis on the particle and the first principal axis.  Finally, throughout this discussion, we have assumed that $\theta_{tr}=90^o$.  If that were not the case, in all pre-factors, the radius $a$ should be replaced with ${a} sin \theta _{tr}$. \\
\indent This expression has important consequences for particles in varying curvature fields. We have focused on constant mean curvature interfaces, for which we define the curvature capillary energy at a given position of the interface $E_{cc} =\Delta E - \Delta {E_{planar}}$: \begin{align}
E_{cc}=- \gamma \pi a^2( \frac{h_{qp}\Delta c_0}{2}\cos 2 \alpha).\label{energy-221}
\end{align}
This predicts a local torque enforcing alignment along the principal axes and a force on the particle propelling it toward high curvature regions. This expression is similar to Equation \ref{pairint}, in which one particle moves in the curvature field made by its neighbor.  In that case, the neighboring particle made a deviatoric curvature field in the interface $\Delta c_0=12 h_{qp}\frac{a^2}{r_{12}^4}cos 2(\phi+ \alpha_1).$
\subsection{Analogies to Electrostatics}
Capillary interactions are similar  to electrostatic interactions \cite{Griffiths}. The electrostatic energy $U$ is often likened to the capillary energy. We explore this concept further for a disk that is attached to a curved  interface. First we compare a disk with a circular, pinned contact line to a grounded disk in an external field. Then we consider a disk with a quadrupolar undulated contact line in comparison with a disk having a quadrupolar edge potential. 

The interface height above the plane of the interface $h$ is analogous to the electrostatic potential $\psi^{out}$ external to the particle. The potential at the particle edge corresponds to boundary conditions on $h$ at the contact line. 
Particles with finite edge potentials, however, have two quantities that are absent in capillarity; these include the electrostatic potential inside the particle $\psi^{in}$  and the charge density at the contact line $\sigma$. 

We evaluate $U$ for a disk in an external field of form 
\begin{align}
{\psi _{ext}} = {\psi _0}\;\frac{r^2}{a^2}\cos 2\phi \label{ext-1}.
\end{align}
\subsubsection{A grounded disk in an external field}\hfill\\
Consider a grounded disk with radius $a$ in an unbounded domain $I$ in a far field potential ${\psi _{ext}}$.  The  potentials inside $\psi^{in}$ and outside of the disk $\psi^{out}$ are: 
\begin{align}
& \psi^{out} = {\psi _0}(\frac{r^2}{a^2} - \frac{{{a^2}}}{{{r^2}}})\cos 2\phi ,\\
&\psi^{in} = 0.
\end{align}
The total electrical energy $U$ is:\\
\begin{align}
U= \frac{1}{2}\!\! \mathop{{\int\!\!\!\!\!\int}\mkern-21mu }\limits_{I+P} {  \;\;\; {\rho ({\bf{r}})\psi ({\bf{r}})} dA}
\end{align}
where $I+P$ is the entire domain, $dA$ is an area element, and $\rho({\bf{r}})$ is the charge density in the system. However, the sole charge is $\sigma$, and the induced charge on the surface of the disk,
\begin{align}
\sigma=\left.{ \bf{e}_r\cdot( - \epsilon  \nabla \psi (r \ge a))} \right|_{r = a}= - 4{\epsilon}{\psi _0}a^{-1}\cos 2\phi,
\end{align}
where ${\bf{e}}_r$ is the unit normal pointing away from the disk, and $\epsilon $ is the relative permittivity of the disk to that of free space. The expression for $U$ can be recast and evaluated:
\begin{align}
U = \frac{1}{2}\int_a^\infty\!\! \int_0^{2\pi }\!\!\!\!\! {\sigma}\delta (r - a)\psi ({\bf{r}})rd\phi dr = 0,
\end{align}
where $\delta$ is the Dirac delta function. Recalling that, for a grounded disk, $\psi(r=a)=0$, $U$ is zero. This result agrees with the capillary energy we have derived for circular, pinned contact lines on curved interfaces, for which the capillary energy is zero.  In particular, there are no terms in $\psi_0^2$, or analogously $(a^2 \Delta c_0)^2$ whose contributions have been the subject of discussion \cite{Yaodisk, Nimasphere, Reply}. 
\subsubsection{A disk with an edge potential}\hfill\\
Consider a dielectric disk of radius $a$ with edge potential $\psi (r = a) = q\cos 2\phi$ in an unbounded domain with far field potential $\psi_{ext}$.  Here, $\psi^{in}$ is finite, as the disk polarizes owing to the edge potential. This finite potential has no analogy in the capillary problem; this will propagate throughout the calculation of $U$. For simplicity, we consider a disk and external domain of the same relative permittivity. The electric potentials inside and outside the disk are subject to the boundary conditions: 
\begin{align}
&{\left. {{\psi ^{in}}} \right|_{r = a}} = {\left. {{\psi ^{out}}} \right|_{r = a}},\label{eq-pot}\\
&{\left. {{{\bf{e}}_r} \cdot (\nabla {\psi ^{in}} - \nabla {\psi ^{out}})} \right|_{r = a}} = \frac{{{\sigma}}}{{{\epsilon}}},\label{maxwel}
\end{align}
\\
with solutions: 
\begin{align}
&{\psi ^{in}} = q\frac{r^2}{a^2}\cos 2\phi,\\
&{\psi ^{out}} = q \frac{a^2}{r^2}\cos 2\phi+{\psi _0}(\frac{r^2}{a^{2}} - \frac{{{a^2}}}{{{r^2}}})\cos 2\phi.
\end{align}
Notice that  $\psi^{out}$ has the same form as $h$ for a particle with a pinned, quadrupolar contact line on a curved interface. The corresponding egde charge density is: 
\begin{align}
{\sigma} = 4\epsilon\frac{(q- \psi _0)}{a}\cos 2\phi.
\end{align}
Since this is the sole charge in this system, $U$ is: 
\begin{align}
U & = \frac{1}{2}\int_0^\infty {\int_0^{2\pi } {{\sigma}\delta (r - a)\psi ({\bf{r}})rd\phi dr} } \nonumber\\
&= - 2\pi {\epsilon}{\psi _0}{q}+\pi {\epsilon}q^2+\pi {\epsilon}q^2.
\end{align}
We break this  expression into three terms; the first two are analogous to the net capillary energy for a particle on a curved interface; these include the curvature capillary energy $E_{cc}$ in Equation (\ref{energy-221}) and the self energy owing to the particle distortion in Equation  (\ref{p1}).  The third term, of the same form as the self energy, is not present in the capillary energy.   
To understand the origin of this term, we use Gauss's law to recast $U$:
\begin{align}
&
&U =  - \frac{{{\epsilon}}}{2}\oint\limits_{\partial (I+P)} {(\psi \nabla \psi )\cdot{\bf{n}}\;dl}  + \frac{{{\epsilon }}}{2}\mathop{{\int\!\!\!\!\!\int}\mkern-21mu }\limits_{I+P} 
 {{{(\nabla \psi )}^2}dA},\label{en-elc-1}
\end{align}
where the second integral can be decomposed into the domains inside and outside of the disk:
\begin{align}
\mathop{{\int\!\!\!\!\!\int}\mkern-21mu }\limits_{I+P} 
 {{{(\nabla \psi )}^2}dA}  = \mathop{{\int\!\!\!\!\!\int} }\limits_P 
 {{{(\nabla {\psi ^{in}})}^2}dA}  + \mathop{{\int\!\!\!\!\!\int} }\limits_{I} 
 {{{(\nabla {\psi ^{out}})}^2}dA}.\label{decomp-v-1}
\end{align} 
Then the contribution to $U$ integrated over $P$ is:
\begin{align}
\frac{\epsilon }{2} \mathop{{\int\!\!\!\!\!\int}}\limits_P 
 {{(\nabla {\psi ^{in}})}^2}dA = \epsilon \pi \int_0^a {\frac{4q^2}{a^4}{r^3}dr}  = \epsilon \pi   q^2.
\end{align}
This is the energy required to polarize the disk. This has no analogy in the capillary problem. The process of charging the particle, which generates the potential inside of the particle, differs from the process of undulating the contact line, which relies on wetting energies or pinning sites.  When this term is disregarded, the analogy between electrostatics and capillarity holds.
\begin{figure}[h]
\includegraphics[scale=0.7]{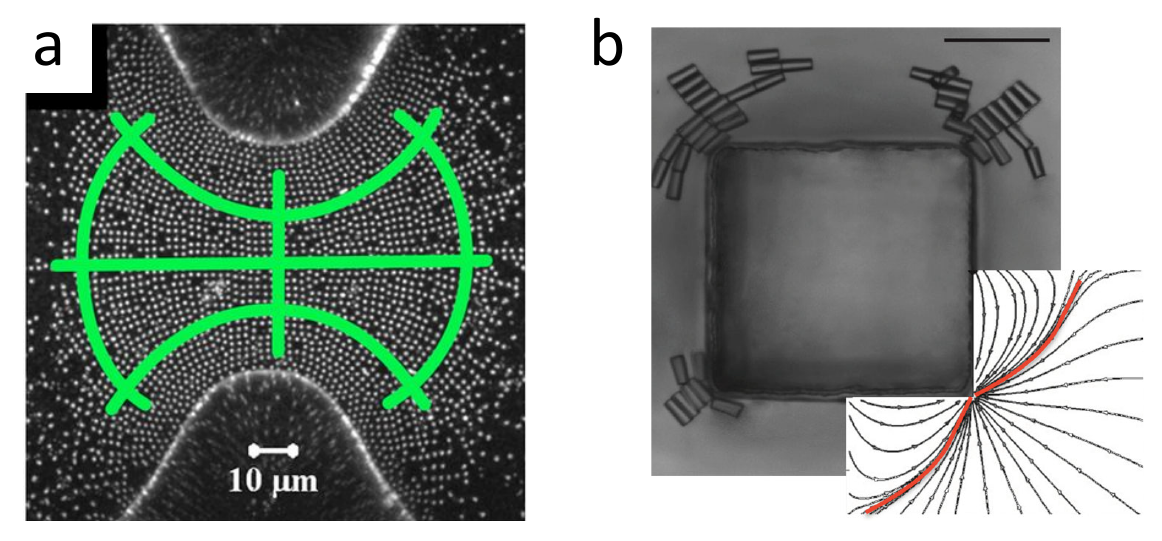}
\caption{Assemblies of particles on curved interfaces. (a) Repulsive microparticles on surface with a negative Gaussian curvature. Reprinted with permission from Reference \cite{Ershov}, copyright (2013), National Academy of Sciences. (b) SU-8 cylindrical particles move in deviatoric curvature gradient to assemble around a square micropost. The red curve demonstrates a ridge of local maxima, along which the structure forms. Scale bar: 100 $\mu m$ Reprinted with permission from Reference \cite{Cavallaro27122011}, copyright (2011), National Academy of Sciences.}
\label{curved_int}
\end{figure}\\ 
\subsection{Key Findings on Curved Interfaces}
\begin{figure}
\includegraphics[scale=0.7]{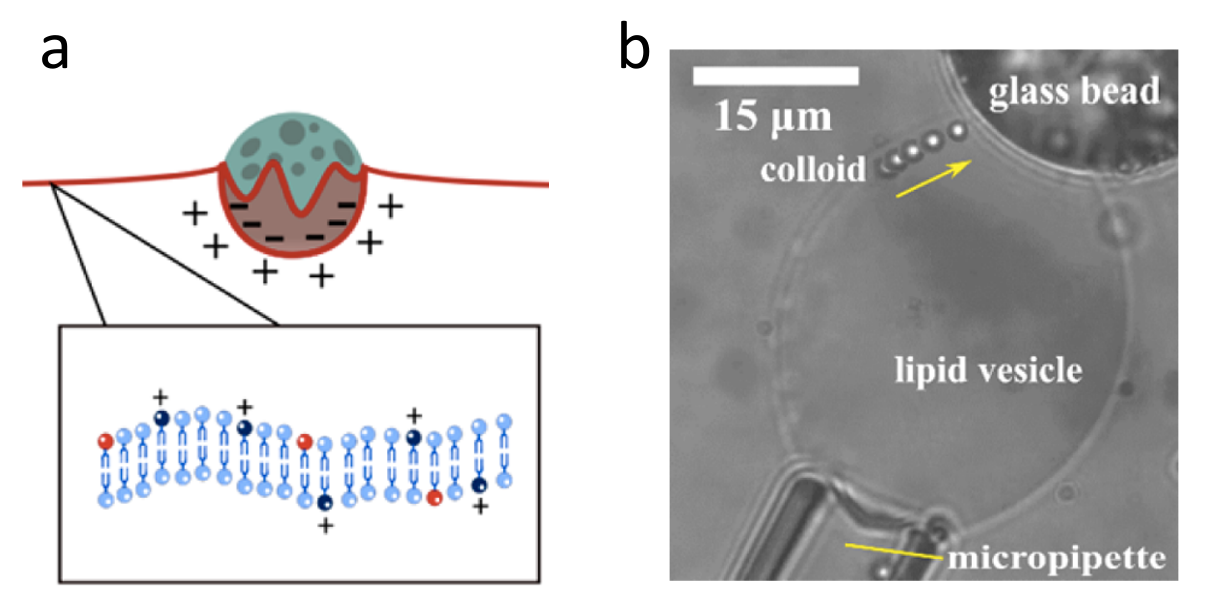}
\caption{Particle trapped on a lipid vesicle. (a) Schematic of a Janus particle wrapped on a lipid bilayer. (b) Particle migration on a tense lipid vesicle. Adapted with permission from \cite{Ningwei}, copyright 2017, ACS. }
\label{lipid}
\end{figure}

Spherical particles on curved interfaces have been analyzed in several limits. Particles with a contact angle of $90 ^\circ$ on bounded cylindrically-shaped interfaces are predicted to induce quadrupolar distortions that drive assembly \cite{ZengSM2012}. Particles with contact lines at equilibrium on unbounded interfaces are predicted either to interact with capillary energies that are quadratic in the deviatoric curvature \cite{Wurger} or to fail to interact \cite{Nimasphere}, depending on the treatment of the far field boundary in evaluating the interfacial area. Since contact line pinning is ubiquitous, this case is difficult to interrogate in experiment. Finally, particles with pinned contact lines on unbounded interfaces have been addressed. This case, which is identical to the case of a disk on the interface except for the constant wetting energy terms, is discussed in section \ref{curved-interface}. \\
\indent Structures that clearly reveal quadrupolar symmetries were observed on interfaces with complex curvatures. The interfaces were formed by placing droplets of oil on surfaces with patterned hydrophobic surfaces; the resulting interfaces had constant mean curvature but spatially varying deviatoric curvatures \cite{Ershov}.  Charged spherical microparticles on these interfaces form a square lattice characteristic of capillary quadrupolar interactions balanced by electrostatic repulsion as shown in Figure \ref{curved_int}a. The square lattice is strained, consistent with the particle quadrupolar distortion aligning along the spatially varying principle axes. At particle densities greater than $0.33$, however, hexagonal lattices begin to appear, indicating that at dense packings, the quadrupolar modes are no longer dominant.
\begin{figure*}
\includegraphics[scale=0.65]{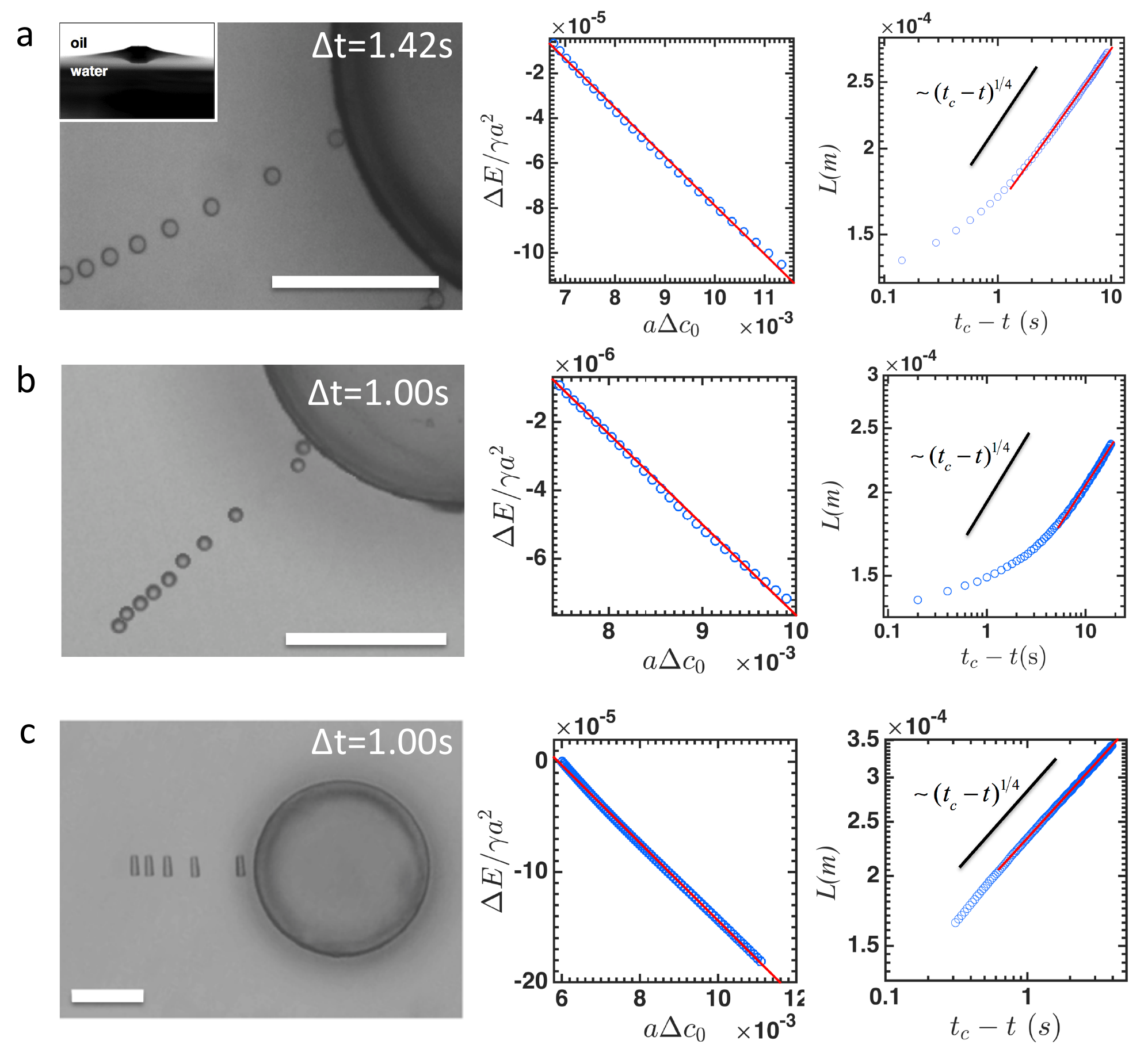}
\caption{Curvature capillary migration of disk (a), sphere (b) and cylinder (c). Left-most column shows the time-stamped image of trajectories of microparticles migrating toward a circular micropost. Center column shows curvature capillary energy for particles are linear in deviatoric curvature. Right-most column plots power law dependence of center-to-center separation distance (L) versus contact time. (a) Inset: side view of the curved interface. Scale bars are 100 $\mu m$. (a) Reprinted from Reference \cite{Yaodisk}, copyright (2015), with permission from Elsevier. (b) Adapted from \cite{Nimasphere} with permission from The Royal Society of Chemistry. (c) Reprinted with permission from \cite{Cavallaro27122011}, copyright (2011), National Academy of Sciences.}
\label{dsc}
\end{figure*}
\indent Spherical particles also respond dynamically to curved interfaces. Silica colloids at air-water interfaces displayed different behaviors depending on particle wetting \cite{Blanc2013PRL}. For contact angles up to $5 ^\circ$, particles settle to the minimum interface height, which is consistent with weak capillary interaction. For a contact angle of $\sim 30 ^\circ$, particles migrate along the principal axis to equilibrium sites.\\
\indent We have studied the dynamics of microparticles on curved water-oil interfaces for several particle shapes including disks \cite{Yaodisk}, spheres \cite{Nimasphere} and cylinders \cite{Cavallaro27122011}. These experiments are performed on interface shapes molded around a micropost. A vertical cylindrical micropost hundreds of microns in diameter is fabricated from epoxy resin on a silicon substrate. The post is surrounded by a low ring, located several capillary lengths away. This space is filled with water so that the contact line pinned is at the edge of the micropost. The maximum slope of the interface occurs at the micropost given by the angle $\psi \sim 15^ \circ$ with the horizontal. A layer of hexadecane is gently placed over this water layer to prevent evaporation, dampen stray convection, and to allow particles to be introduced to the interface via sedimentation through the oil. This apparatus is placed under an optical microscope and imaged from above; particle trajectories are recorded and analyzed. Near the post, in the region of interest, the interface height decays logarithmically with distance $L$ from the center of the micropost. In this region, the mean curvature is negligible and the deviatoric curvature $\Delta {{c}_{0}}$ is known, $\sim -L^{-2}$. Particles attach to the interface and migrate along the curvature gradient to sites of high deviatoric curvature, in agreement with the predictions in Equation \ref{energy-221}. Results are summarized in Figure \ref{dsc}, including time-stamped images of typical trajectories, with the energy dissipated along a particle path and observed power laws implied by the equality of viscous drag and the curvature capillary force. The energy dissipated versus deviatoric curvature is linear in $\Delta {{c}_{0}}$ with coefficient of linear regression ${R^2} = 0.999$ or better. As the cylindrical microparticle migrates, it aligns its quadrupolar rise axis with the rise axis toward the micropost, in agreement with the capillary torque implied by Equation \ref{energy-221}. Recent work reveals analogies between capillary migration on curved fluid interfaces and Janus bead migration on tense lipid bilayer vesicles (Fig.\ref{lipid}).  The tense vesicle shape obeys the Young-Laplace equation.  The particle migrates along curvature gradients on GUVs, giant unilamellar vesicles, stretched to impose curvature fields, with energy dissipated linearly in the deviatoric curvature \cite{Ningwei}.  For all of these studies, the interface slope is small and the particle is much smaller than the micropost. In recent work, we have challenged these assumptions. That discussion is outside of the scope of this review. 
\subsection{Summary}
\indent The discussion of lattice formation by repulsive particles on interfaces with weakly varying deviatoric curvatures is one interesting limit for structure formation on curved interfaces. Particles formed a square lattice, and neither assembled nor migrated. Structure formation on interfaces with strong curvature gradients is another interesting limit. For example, consider cylindrical particles on an interface pinned to a square micropost (see Fig. \ref{curved_int}b). Curvature gradients near the corners are very steep; particles migrate along curvature gradient lines to form trapped structures influenced by the curvature gradients, and particle pair interactions, and potentially by multibody effects and local non-linearities. Between these limits, tunable structures informed by curvature gradients may be formed. This is a focus of ongoing work.  
\section{Conclusions}
Capillary interactions between colloids at fluid interfaces direct their assembly into structures with preferred orientations, cemented by capillary bonds whose strength depends strongly on particle shape, including features like elongations, the existence of facets or sharp edges and roughness. Contact line pinning plays a major role; contact line undulations on the scale of nanometers can create significant interface distortions and capillary interactions. Interface curvature-in particular, finite deviatoric curvature-plays a central role in assembly and in guiding structure formation. In this review, we have summarized the key arguments behind pair interactions and curvature capillary interactions for interfaces with small slopes. We have endeavored to summarize several findings that exemplify different aspects of the relationship between particle shape, contact line pinning, interface shape and interactions. Work in the field is in its early stages and developing at a rapid pace, both in terms of directed assembly of particles, and in terms of fundamental understanding of the interactions and their implications in other, related systems.

\section*{DISCLOSURE STATEMENT}
The authors are not aware of any affiliations, memberships, funding, or financial holdings that might be perceived as affecting the objectivity of this review. 

\section*{ACKNOWLEDGMENTS}
This work is supported by NSF DMR-1607878.

%



\bibliographystyle{ar-style4.bst}
%
\newcommand{\noopsort}[1]{} \newcommand{\printfirst}[2]{#1}
  \newcommand{\singleletter}[1]{#1} \newcommand{\switchargs}[2]{#2#1}

\end{document}